\documentclass[aps,pra,reprint,superscriptaddress]{revtex4-1}

\usepackage{graphicx,color}
\usepackage{amsmath}
\usepackage{dcolumn}
\usepackage{bm}
\usepackage{hyperref}
\usepackage{braket}
\usepackage{ulem}

\newcommand{\be}{\begin{eqnarray}}
\newcommand{\ee}{\end{eqnarray}}

\graphicspath{{./images/}}

\begin{document}

\title{Semiclassical dynamics of a dark soliton in a one-dimensional bosonic superfluid in an optical lattice}

\author{Yusuke Ozaki}
\email[Electronic address : ]{Y.OZAKI@kindai.ac.jp}
\affiliation{Department of Physics, Kindai University, Higashi-Osaka, Osaka 577-8502, Japan}
\author{Kazuma Nagao}
\affiliation{Institut f\"ur Laser-Physik and Zentrum f\"ur Optische Quantentechnologien, Universit\"at Hamburg, D-22761 Hamburg, Germany}
\affiliation{The Hamburg Center of Ultrafast Imaging, Luruper Chaussee 149, D-22761 Hamburg, Germany}
\author{Ippei Danshita}
\email[Electronic address : ]{danshita@phys.kindai.ac.jp}
\affiliation{Department of Physics, Kindai University, Higashi-Osaka, Osaka 577-8502, Japan}
\author{Kenichi Kasamatsu}
\affiliation{Department of Physics, Kindai University, Higashi-Osaka, Osaka 577-8502, Japan}

\date{\today}

\begin{abstract}
    We study quantum dynamics of a dark soliton in a one-dimensional Bose gas in an optical lattice within the truncated Wigner approximation.
    A previous work has revealed that in the absence of quantum fluctuations, dynamical stability of the dark soliton significantly depends on whether its phase kink is located at a lattice site or a link of two neighboring sites. 
    It has also shown that the dark soliton is unstable in a regime of strong quantum fluctuations regardless of the phase-kink position. 
    To bridge the gap between the classical and strongly quantum regimes, we investigate the dynamical stability of the dark soliton in a regime of weak quantum fluctuations. 
    We find that the position dependence of the dynamical stability gradually diminishes and eventually vanishes as the strength of quantum fluctuations increases.
    This classical-to-quantum crossover of the soliton stability remains even in the presence of a parabolic trapping potential. 
    We suggest that the crossover behavior can be used for experimentally diagnosing whether the instability of a dark soliton is due to quantum fluctuations or classical dynamical instability.
    \end{abstract}

\maketitle


\section{Introduction} \label{ss1_introduction}
Solitons are robust nonlinear waves, which are localized and collide elastically with one another like particles. 
Soliton solutions in classical systems are described as analytical solutions of integrable nonlinear equations, including the Korteweg-de Vries equation~\cite{zabusky_interaction_1965}, the sine-Gordon equation~\cite{mandelstam_soliton_1975}, and nonlinear Schr\"odinger equations~\cite{kivshar_dark_1998}.
The nonlinear Schr\"odinger equation with cubic nonlinearity, namely the Gross-Pitaevskii (GP) equation, can describe Bose-Einstein condensates (BECs) of ultracold atomic gases in a weakly interacting regime~\cite{dalfovo_theory_1999}.
Experiments with ultracold atoms are suited for studying dynamical processes of BECs in an ideal situation because the system is well isolated from environment and the relaxation time is sufficiently longer than the typical time scale of interesting dynamical phenomena.
Taking these advantages, previous experiments have extensively investigated dynamical properties of solitons in atomic BECs~\cite{burger_dark_1999, denschlag_generating_2000, khaykovich_formation_2002, becker_oscillations_2008, PhysRevLett.101.130401, Aycock2503}.

Previous theoretical studies have revealed that a dark soliton state of a BEC is dynamically stable when the BEC is a one-dimensional (1D) continuous system~\cite{tsuzuki_nonlinear_1971, muryshev_stability_1999}.
Recent experiments~\cite{gring_relaxation_2012, langen_experimental_2015, schweigler_experimental_2017, erne_universal_2018} have realized highly 1D BECs, whose radial size is much smaller than the healing length, enabling exploration of quantum phenomena in 1D systems.
Such experimental developments in creating 1D BECs indicate that a stable dark soliton can be experimentally observed in the near future.
Since the strength of quantum fluctuations is widely controllable in the system of a 1D BEC by tuning the density, the interatomic interaction,  or  the depth of an optical lattice, previous works have investigated the effects of quantum fluctuations on a dark soliton by means of various theoretical methods, such as the Bogoliubov theory~\cite{dziarmaga_quantum_2004}, the Bethe ansatz~\cite{sato_exact_2012}, the matrix product states (MPS)~\cite{mishmash_quantum_2009, mishmash_quantum_2009-2, delande_many-body_2014}, and the truncated Wigner approximation (TWA)~\cite{martin_quantum_2010, martin_nonequilibrium_2010-1}.
In particular, Mishmash {\it et al.}~have analyzed stability of a dark soliton in the presence of an optical lattice in the classical regime by the Bogoliubov approximation and in a regime of strong quantum fluctuations by using the MPS~\cite{mishmash_quantum_2009}.
There is a remarkable difference in the classical dynamical stability depending on whether the phase kink of the dark soliton is located at a lattice site or at a link of two neighboring sites~\cite{PhysRevE.75.066608}.
They have also revealed that by contrast strong quantum fluctuations destabilize a dark soliton regardless of its location.
Hence, one can naturally expect that the stability of the dark soliton changes between the classical regime and the regime of strong quantum fluctuations.

In this paper, we study the dynamical stability of the dark soliton of a 1D lattice Bose gas in a regime of weak quantum fluctuations.
First, we reconsider the stability of the dark soliton in the classical limit by solving the Bogoliubov equations numerically. 
The critical interaction strength, above which the soliton state is dynamically unstable, is determined from the condition for a complex normal-mode frequency to emerge. 
We discuss reasons why the soliton stability depends on its kink position by considering the difference in the energy gain of the system due to a small-amplitude fluctuation.
Second, we numerically analyze dynamics of the dark soliton including quantum fluctuations within the TWA~\cite{HILLERY1984121, blakie_dynamics_2008, polkovnikov_phase_2010}. 
We find that even in the semiclassical regime, quantum fluctuations significantly affect the stability of the dark soliton.
Specifically, when the strength of quantum fluctuations increases, the distinction in the soliton stability with respect to the kink position gradually disappears.
We show that this disappearance can be attributed to the fact that stronger fluctuations cause a soliton oscillation with larger amplitude regardless of the kink position, thus invaliding the classical stability analysis.

This paper is organized as follows.
In Sec.~\ref{ss2_model}, we introduce the Bose-Hubbard model, which describes ultracold bosons in a deep optical lattice.
In Sec.~\ref{ss3_methods}, we review theoretical methods used for analyzing stability of a dark soliton.
Specifically, in Sec.~\ref{ss3_sssa_bogo}, we formulate a linear stability analysis of BECs in the classical limit on the basis of the GP mean-field theory.
In Sec.~\ref{ss3_sssb_twa}, we explain the TWA, which allows for computation of quantum dynamics of the Bose-Hubbard model within the semiclassical regime.  
In Sec.~\ref{ss4_sssa_box_cla}, we present results of the linear stability analysis on a dark soliton in a box potential in a wide range of parameters and discuss reasons why the soliton stability significantly depends of its kink position.
In Sec.~\ref{ss4_sssb_box_weak}, we analyze the semiclassical dynamics of the dark soliton in a box potential by means of the TWA.
In Sec.~\ref{ss5_sssa_para_cla} and~\ref{ss5_sssb_para_weak}, we perform the numerical analyses in the case that a parabolic trapping potential is present. 
In Sec.~\ref{ss6_conclusion}, we summarize our results.


\section{Model} \label{ss2_model}
We consider an ultracold Bose gas elongated in one spatial direction. 
We assume that the confinement in the transverse directions is sufficiently strong so that thermal and dynamical excitations in the transverse directions can be safely neglected, i.e., the system is 1D. 
Such strong confinement can be created with use of optical lattices \cite{moritz_exciting_2003, kinoshita_observation_2004} or atom chips~\cite{gring_relaxation_2012, langen_experimental_2015, schweigler_experimental_2017, erne_universal_2018}. 
We also assume that an optical lattice is present in the longitudinal direction and is sufficiently deep. 
Then, our system can be described by the Bose-Hubbard Hamiltonian (BHH) \cite{fisher_boson_1989, jaksch_cold_1998}.
\begin{align}
\begin{aligned}
 \hat{H}=&-J\sum_{j=1}^{M-1}(\hat{b}_{j+1}^{\dag}\hat{b}_j+{\rm h.c.})  \\
 &+\frac{U}{2}\sum_{j=1}^{M}\hat{n}_j(\hat{n}_j-1)+\sum_{j=1}^{M}\epsilon_{j}\hat{n}_j,
 \end{aligned}
 \label{eq:BHH}
 \end{align}
 where $J$ is the nearest-neighbor hopping amplitude, $U(>0)$ is the on-site interaction coefficient, $\epsilon_{j}$ is an external potential, $M$ is the total number of sites, $\hat{b}_{j}$ and $\hat{b}_{j}^{\dag}$ are the bosonic annihilation and creation operators at $j$-th site, and the number operator {$\hat{n}_j=\hat{b}_{j}^{\dag} \hat{b}_{j}$} counts the number of particles at $j$-th site.
 The summation over index $j$ on the hopping term contains only $M-1$ terms under open (box) boundary conditions.
 
The ground-state phase diagram of the BHH includes the superfluid and Mott insulator phases, which are separated by a continuous quantum phase transition \cite{fisher_boson_1989}. 
When the dimensionless parameter, $\gamma = U/(z\nu J)$, is small, the system favors the superfluid phase, where $z$ and $\nu$ denote the coordination number and the filling factor, respectively. 
When $\nu$ is integer, the Mott insulator phase is favored at large $\gamma$. 
The quantum phase transition across the two phases has been indeed observed in experiments with ultracold bosons in optical lattices in 1D \cite{stoferle_transition_2004}, 2D \cite{spielman_mott-insulator_2007}, and 3D \cite{greiner_quantum_2002}.
The critical value of the quantum phase transition at unit filling in 1D has been obtained as $(U/J)_c=3.367$ with numerical methods \cite{kuhner_one-dimensional_2000}.
The critical values of the dimensionless parameter, $\gamma_c$, for the transitions in 1D, 2D, and 3D have been precisely determined for arbitrary filling factors in previous literature~\cite{PhysRevB.53.2691, PhysRevB.58.R14741, kuhner_one-dimensional_2000, PhysRevB.75.134302, PhysRevA.77.015602, PhysRevB.79.100503, PhysRevB.79.224515, doi:10.1063/1.3046265, Ejima_2011, danshita_superfluid--mott-insulator_2011}.
Throughout the present work, we assume that the system is deep in a superfluid region, i.e., $\gamma \ll 1$, where the semiclassical TWA can describe its quantum dynamics at least qualitatively. 


\section{Methods} \label{ss3_methods}

\subsection{Linear stability analysis in the classical limit} \label{ss3_sssa_bogo}
In this section, we review the GP mean-field theory for the BHH, in which dynamics of a BEC is described by the discrete nonlinear Schr\"odinger equation (DNLSE)~\cite{amico_dynamical_1998, trombettoni_discrete_2001}.
We explain how to analyze stability of stationary solutions of the DNLSE on the basis of the Bogoliubov equations~\cite{pitajevskii_bose-einstein_2003}.

The Heisenberg equation of motion for the annihilation operator $\hat{b}_j$ is given by
\begin{eqnarray}
i\hbar \frac{\partial}{\partial t} \hat{b}_j=\bigl[\hat{b}_j,\hat{H}\bigr].
\label{eq:heis}
\end{eqnarray}
By replacing $\hat{b}_j$ with its mean field, $\langle\hat{b}_j\rangle\equiv\Psi_j$, under the assumption that the quantum and thermal  depletions are negligibly small, we obtain the DNLSE,
 \begin{equation}
i\hbar\frac{\partial }{\partial t}\Psi_{j}=-J(\Psi_{j+1}+\Psi_{j-1})+U|\Psi_j|^2\Psi_j+\epsilon_{j} \Psi_j.
\label{eq:tdDNLSE}
\end{equation}

Let us seek a solution of the DNLSE describing small fluctuations around a stationary solution $\Phi_j$,
\begin{equation}
 \Psi_j(t)=e^{-i\mu t/\hbar}\bigl[\Phi_j+\delta\Psi_j(t)\bigr],
\end{equation}
where $\mu$ is the chemical potential and $\Phi_j$ obeys the time-independent DNLSE,
\begin{eqnarray}
\mu \Phi_{j}=-J(\Phi_{j+1}+\Phi_{j-1})+U|\Phi_j|^2\Phi_j+\epsilon_{j} \Phi_j.
\label{eq:tiDNLSE}
\end{eqnarray}
The fluctuation part, $\delta \Psi_j(t)$, can be expanded as the following form:
\begin{equation}
\centering
\delta \Psi_j(t)=\sum_{\ell}\left[ u_j^{(\ell)} e^{-i\omega_{\ell} t}-v^{(\ell) \ast}_j e^{i\omega_{\ell}^{\ast} t} \right],
\label{eq:yrg}
\end{equation}
where $u_j^{(\ell)}$ and $v_j^{(\ell)}$ are the amplitudes of the $\ell$-th normal mode and $\omega_{\ell}$ is the associated mode frequency.
By substituting Eq.~(\ref{eq:yrg}) into Eq.~(\ref{eq:tdDNLSE}), we obtain the following pair of coupled equations for $u_j^{(\ell)}$ and $v_j^{(\ell)}$, namely the Bogoliubov equations:
\begin{align}
    \begin{aligned}
-J(u^{(\ell)}_{j+1}+u^{(\ell)}_{j-1})+2U|\Phi_j|^2 u^{(\ell)}_j+\epsilon_j u^{(\ell)}_j\;\;\;\;\;\;\;\;& \\
-U \Phi^2_j v^{(\ell)}_j =\hbar \omega_{\ell} u^{(\ell)}_j&, 
    \end{aligned}\\
    \begin{aligned}
J(v^{(\ell)}_{j+1}+v^{(\ell)}_{j-1})-2U|\Phi_j|^2 v^{(\ell)}_j-\epsilon_j v^{(\ell)}_j\;\;\;\;\;\;\;\;& \\
+U {\Phi_j^\ast}^2 {u_j^{(\ell)}} =\hbar \omega_{\ell} v^{(\ell)}_j. &
\end{aligned}
\end{align}
Solving these simultaneous equations, we obtain $2M$ eigenvectors $(u^{(\ell)}_j, v^{(\ell)}_j)$ and eigenvalues $\omega_{\ell}$. 

The stability of the stationary solution $\Phi_j$ against the small fluctuations can be identified with the eigenvalues and eigenvectors of the Bogoliubov equations \cite{pitajevskii_bose-einstein_2003}.
If there exists a normal mode whose ${\rm Im}[\omega_{\ell}]$ is finite, the stationary solution $\Phi_j$ is dynamically unstable.
If all the normal-mode frequencies are real and there exists a normal mode with negative $\omega_{\ell}$ and positive norm, $\sum_{j} [|u_j^{(\ell)}|^2 -|v_j^{(\ell)}|^2]>0$, the stationary solution $\Phi_j$ is energetically unstable.
However, as long as the system is well decoupled from the environment as in typical ultracold-gas experiments, the energetic instability, called Landau instability, does not make the system unstable at sufficiently low temperatures, where the thermal depletion from the BEC is negligibly small \cite{PhysRevA.72.013603}.
Other than these two cases, the stationary solution is stable.

In Sec.~\ref{ss4_sssa_box_cla}, we use the formulation described above in order to analyze the stability of stationary dark-soliton solutions.
To obtain the stationary solutions, we perform the imaginary-time propagation of Eq.~(\ref{eq:tdDNLSE}) under the constraint condition that the phase of $\Phi_j$ jumps by $\pi$ across the center of the system.


\subsection{Truncated Wigner approximation} \label{ss3_sssb_twa}
In this section, we briefly review the TWA method applied to the BHH~\cite{HILLERY1984121, blakie_dynamics_2008, polkovnikov_phase_2010}.
A quantum state can be in general described by a distribution function in a classical phase space.
With the phase space method, the time evolution of quantized particles or fields can be expressed by the Wigner function $W(\vec{\alpha}, \vec{\alpha}^\ast, t)$, which corresponds to a quasi-probability distribution and the Wigner-Weyl transform of the density matrix $\hat{\rho}(t)$.
In the TWA for BHH, we consider a 2$M$-dimensional phase space of a complex-valued vector $\vec{\alpha}=(\alpha_1, \alpha_2, \cdots \alpha_M)$.
The time evolution of the Wigner function can be obtained by performing the Wigner-Weyl transform of the von-Neumann equation as
\begin{equation}
i \hbar \frac{\partial}{\partial t} W(\vec{\alpha}, \vec{\alpha}^\ast, t)=2H_W(\vec{\alpha}, \vec{\alpha}^\ast) \sinh\left(\frac{\Lambda_c}{2}\right) W(\vec{\alpha}, \vec{\alpha}^\ast, t),
\label{eq:VNeq}
\end{equation}
where $H_W=(\hat{H})_W$ is the Wigner-Weyl transform of Hamiltonian, $\Lambda_c=\sum_j \left[\overleftarrow{\partial}_{\alpha_j} \overrightarrow{\partial}_{\alpha_j^\ast}- \overleftarrow{\partial}_{\alpha_j^\ast} \overrightarrow{\partial}_{\alpha_j}\right]$ is the symplectic operator acting on $c$-number functions defined in the phase space~\cite{blakie_dynamics_2008, polkovnikov_phase_2010}.
The expectation value of an operator $\hat{A}(t)$, which is expressed by $\langle \hat{A}(t) \rangle ={\rm Tr}[\hat{\rho}(t) \hat{A}]$, can be described as
\begin{equation}
\langle \hat{A}(t) \rangle=\int d\vec{\alpha} d\vec{\alpha}^\ast W(\vec{\alpha}, \vec{\alpha}^\ast, t) A_W(\vec{\alpha}, \vec{\alpha}^\ast),
\end{equation}
where $d\vec{\alpha} d\vec{\alpha}^\ast=\pi^{-M}\prod_j d {\rm Re}[\alpha_j] d{\rm Im}[{\alpha_j}]$, $A_W=(\hat{A})_W$ is the Wigner-Weyl transform of $\hat{A}$~\cite{blakie_dynamics_2008, polkovnikov_phase_2010}.

The complete information about the quantum dynamics can be obtained by solving Eq.~(\ref{eq:VNeq}).
However, it is difficult to solve Eq.~(\ref{eq:VNeq}) exactly.
Instead, we perform a semiclassical expansion of the right hand side of Eq.~(\ref{eq:VNeq}) with respect to the symplectic operator $\Lambda_c$.
If one neglects higher-order terms of $\mathcal{O}(\Lambda^3_{c})$, which is called TWA~\cite{HILLERY1984121, blakie_dynamics_2008, polkovnikov_phase_2010}, then the time evolution of the Wigner function is described by the classical Liouville equation $i \hbar \partial_t W \approx \{H_W, W\}_{\rm P.B.} = H_W \Lambda_{c} W$.
Here the bracket $\{\cdot, \cdot\}_{\rm P.B.}$ represents the Poisson bracket defined in the phase space.
In the TWA, the Wigner function is conserved in time along the classical trajectories, which are solutions of the DNLSE,
\begin{equation}
i \hbar \frac{\partial \alpha_{{\rm cl}, j}}{\partial t}=\frac{\partial H_W}{\partial \alpha^\ast_{cl, j}}.
\end{equation}
By replacing $\alpha_{{\rm cl},j}$ with $\psi_j$, one can write the explicit form of the DNLSE,
 \begin{align}
\begin{aligned}
 i\hbar \frac{\partial \psi_j}{\partial t}=&-J(\psi_{j+1}+\psi_{j-1})\\
 &+U(|\psi_j|^2-1)\psi_j+\epsilon_j \psi_j.
\end{aligned}
\end{align}
The expectation value of an operator $\hat{A}(t)$ can be reduced to a semiclassical form,
\begin{equation}
\langle \hat{A}(t) \rangle \approx \int d\vec{\psi}_0 d\vec{\psi}^\ast_0 W_0\left(\vec{\psi}_0, \vec{\psi}_0^\ast\right) 
A_W\left[\vec{\psi}(t), \vec{\psi}^\ast(t)\right], 
\end{equation}
where $\vec{\psi}(t)=\vec{\psi}\left(\vec{\psi}_0, \vec{\psi}_0^\ast, t\right)$ is a solution of the DNLSE for an initial classical field $\vec{\psi}_0$.
The initial classical fields distribute over the phase space according to the Wigner function of the initial quantum state, $W_0 \equiv W(t=0)$, which describes a leading-order correction of quantum fluctuations to the mean-field solution \cite{polkovnikov_quantum_2003}. 
If $\gamma = 0$ in BHH, corresponding to $U/J=0$ or $\nu=\infty$, the semiclassical approximation becomes exact.
As long as the condition $\gamma \ll 1$ is satisfied, the TWA captures quantum dynamics of the BHH at least qualitatively~\cite{polkovnikov_phase_2010, nagao_semiclassical_2019-1}.

We assume that the system is initially in a direct-product state of the local coherent state labeled by a complex number ${\bar \psi}_{j}$,
\begin{eqnarray}
\ket{\Psi_{\rm ini}}=\bigotimes_{j=1}^{M} e^{ {\bar \psi}_{j}{\hat b}^{\dagger}_{j}- {\bar \psi}_{j}^\ast{\hat b}_{j}}\ket{0},
\label{eq:cohe}
\end{eqnarray}
where $\ket{0}$ is the vacuum of the lattice bosons. Plugging ${\bar \psi}_{j}=\Phi_{j}$ into each coherent state and performing the Wigner-Weyl transform of this pure state, we have the following distribution function as $W_0$~\cite{polkovnikov_phase_2010}:
\begin{eqnarray}
W_0\left(\vec{\psi}_{0},\vec{\psi}_{0}^\ast\right)=\prod_{j=1}^{M}\left\lbrace2\exp\left[-2|\psi_{j,0}-\Phi_j|^2\right]\right\rbrace.
\label{eq:gauss-dist}
\end{eqnarray}
We recall that $\Phi_j$ is a solution of the time-independent DNLSE (\ref{eq:tiDNLSE}), which is a dark-soliton state in the calculations below. 
For all computations of the TWA in real time, we perform Monte-Carlo sampling of 50000 random initial states of the classical fields according to the probability distribution of Eq. (\ref{eq:gauss-dist}) and use the standard fourth-order Runge-Kutta routine for time evolution.

The coherent state of Eq.~(\ref{eq:cohe}) is the ground state of the BHH at $\nu U/J=0$~\cite{polkovnikov_phase_2010}. 
However, we choose this type of state as an initial state for time evolution of a dark soliton at finite $\nu U/J$. 
Our main motivation for this choice is that the Wigner function of the coherent state takes the Gaussian form of Eq.~(\ref{eq:gauss-dist}), which is rather simple from a theoretical viewpoint. 
We aim to address effects of the quantum fluctuations generated by the simple Wigner function on the stability of a dark soliton. 
It is worth noting that while the choice of the coherent state makes it difficult to prepare the same initial state in experiments, it is still feasible with currently available techniques. 
Specifically, one can utilize the Feshbach resonance to set $U=0$ and also can design the external potential $\epsilon_j$, e.g., with use of a digital micromirror device \cite{mazurenko_cold-atom_2017} in such a way that the density profile of the ground state almost coincides with that of a dark-soliton state at finite $\nu U/J$. 
By applying the phase imprinting method~\cite{burger_dark_1999, denschlag_generating_2000, becker_oscillations_2008} to the prepared state and quench $\nu U/J$ towards a desired value, the subsequent dynamics mimics time evolution of a dark soliton at the finite $\nu U/J$ in the presence of quantum fluctuations generated by the initial coherent state, which we will study in the following sections.


\section{System with a box potential} \label{ss4_box_results}

In this section, for theoretical simplicity, we assume that $\epsilon_j$ is homogeneous and the boundary condition is open. 
This means that a Bose gas is confined in a box potential. 
While a trapping potential of parabolic shape is used more often in experiments for confining ultracold gases, some previous experiments have successfully confined ultracold gases in box potentials~\cite{meyrath_bose-einstein_2005, gaunt_bose-einstein_2013, mukherjee_homogeneous_2017}.

\subsection{Classical regime} \label{ss4_sssa_box_cla}
\begin{figure}
\centering
\includegraphics[width=0.9\linewidth] {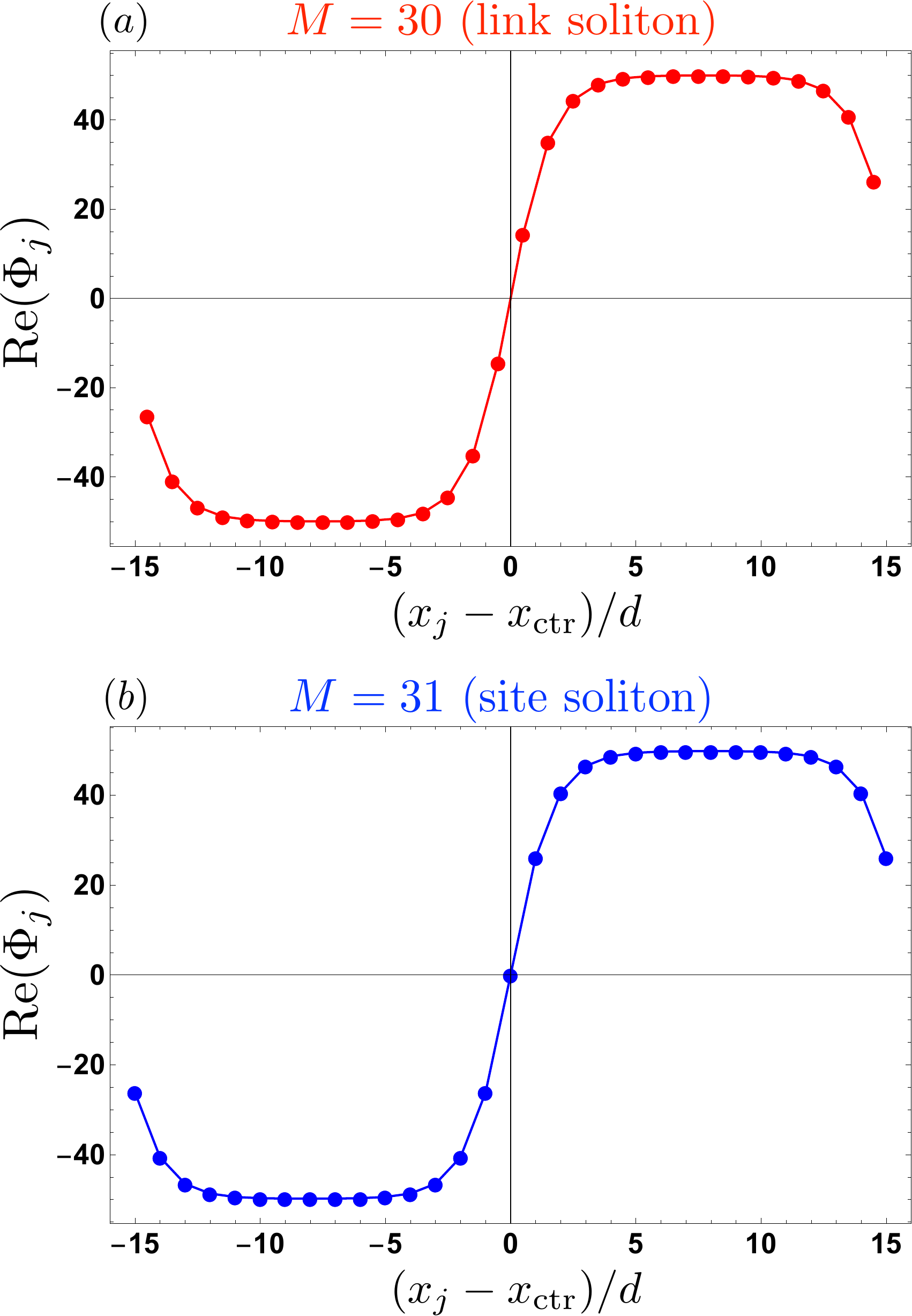}
\caption{The spatial profile of the stationary solutions $\Phi_j$ at $\kappa = 0.5$ and $\nu=2000$ for (a)$M=30$ and (b)$M=31$. The solid lines are guide to the eye. $x_{\rm ctr}=(M+1)/2$ denotes the central position of the system.}
\label{fig_dsols}
\end{figure}

In Fig.~\ref{fig_dsols}, we show the spatial profile of the stationary solutions $\Phi_j$ of the DNLSE at $\kappa = 0.5$ for $M=30$ and $31$ that possess a dark soliton at the center of the system, which are obtained by the imaginary-time propagation of Eq.~(\ref{eq:tdDNLSE}).
Notice that the dimensionless parameter $\kappa \equiv \nu U/J$ quantifies the strength of the nonlinear term relative to the hopping term in the DNLSE~\cite{PhysRevA.66.053607}. 
It is clearly seen that the phase kink of the dark soliton for $M=31$ is located at the central site while that for $M=30$ is located at the link between the two central sites~\cite{PhysRevE.50.5020}. 
For later convenience, we call the former ``a site soliton" and the latter ``a link soliton", respectively. 
Previous studies have shown that the difference in the kink position significantly affects the stability of the dark-soliton solutions in the classical limit~\cite{PhysRevE.75.066608, mishmash_quantum_2009-2}.
In this section, we briefly review the linear stability analysis of Ref.~[\onlinecite{mishmash_quantum_2009-2}] and discuss reasons why the stability of the dark soliton depends on the kink position. We also extend the linear stability analysis to a wider parameter region.

\begin{figure}
\includegraphics[width=0.9\linewidth] {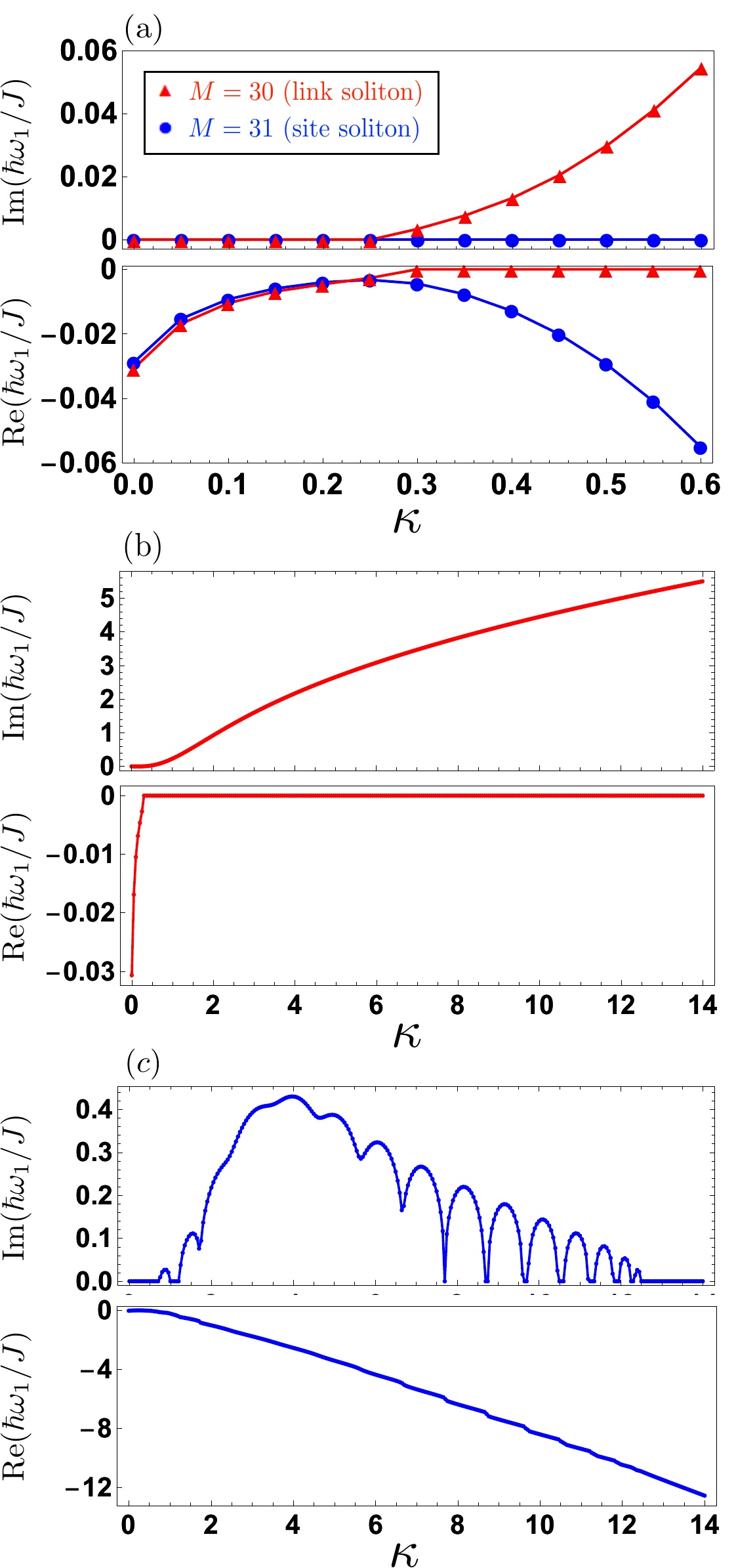}
\vspace{-1mm}
\caption{Frequency $\hbar \omega_1 / J$ of the normal mode that is relevant to the energetic and dynamical instability of the dark soliton as a function of $\kappa$, where $M=30$ (red triangles) and $31$ (blue circles). (a) displays $\hbar \omega_1 / J$ in the small-$\kappa$ region, $0\leq \kappa \leq 0.6$, for both $M=30$ and $31$, which is a reproduction of Fig. 11 of Ref.~[\onlinecite{mishmash_quantum_2009-2}]. (b) and (c) display $\hbar \omega_1 / J$ in the expanded region, $0\leq \kappa \leq 14$, for $M=30$ and $31$, respectively.}
\label{3031_bogo_uni}
\end{figure}
Figure~\ref{3031_bogo_uni} depicts the frequency $\hbar \omega_1 /J$ of the normal mode that is relevant to the energetic and dynamical instabilities of the dark soliton as a function of $\kappa$.
When $\kappa$ is smaller than a certain critical value $\kappa_c \simeq 0.3$, $\hbar \omega_1 /J$ is real and negative for both $M=30$ and $31$. 
This normal mode corresponds to an oscillation of a dark-solition position around the center of the system~\cite{mishmash_quantum_2009-2}.
When $\kappa$ exceeds $\kappa_c$, the real part of $\hbar \omega_1/J$ vanishes and its imaginary part emerges in the case of $M=30$, signaling the dynamical instability of the link soliton. 
By contrast, when $\kappa$ increases above $\kappa_c$ in the case of $M=31$, the real part of $\hbar \omega_1/J$ grows. 
As shown in Fig.~\ref{3031_bogo_uni}(c), when $\kappa$ exceeds $\kappa \simeq 0.7$, the imaginary part of $\hbar \omega_1/J$ starts to grow, signaling the dynamical instability of the site soliton, but behaves rather irregularly. 
This irregular behavior is due to the crossing with anti-modes whose norm is negative, i.e., $\sum_{j} [|u_j^{(\ell)}|^2 -|v_j^{(\ell)}|^2]<0$. 
It is known that such a coupling between a mode and anti-modes results in the emergence of dynamical instability \cite{taylor_bogoliubov_2003-1}. 
Indeed, ${\rm Im}(\hbar\omega_1/J)$ vanishes when $\kappa > 13$, in which $|{\rm Re}(\hbar\omega_1/J)|$ is larger than the frequencies of all the other modes so that there is no anti-mode coupled with the normal mode. 
Thus, the dynamical stability of the dark soliton significantly depends on the location of its kink.

Let us discuss mechanisms how the difference between the link and site solitons with respect to the dynamical stability emerges. 
For this purpose it is useful to pay attention to the three important length scales, namely the lattice spacing $d$, the system size $L = Md$, and the healing length $\xi = d\sqrt{2/\kappa}$. 
In the limit of $\kappa \rightarrow 0$, $\xi$ diverges such that the condition $\xi \gg L \gg d$ holds. In this situation, the dark soliton state reduces to the first excited state of the linear Schr\"odinger equation and the size of its density notch is set by $L$.

In the region of $0.05 \leq \kappa \leq 0.2$, where the condition $L \gg \xi \gg d$ holds, the size of the density notch is comparable to $\xi$. 
There the continuous approximation is safely valid so that the site and link solitons are almost equivalent to each other. In this region, $\hbar \omega_1/J$ approaches zero with increasing $\kappa$ for the following reason. 
In a continuous system, the system in the ground state has approximate translational symmetry and the presence of a dark soliton breaks the symmetry. 
This means that the dark-soliton state has a Nambu-Goldstone mode with $\omega = 0$. 
When $\kappa$ increases, the healing length $\xi$ decreases so that the normal mode with the frequency $\omega_1$ approaches the Nambu-Goldstone mode.

In the region of $\kappa \gtrsim \kappa_c$, $\xi$ is comparable to $d$ so that the continuous approximation breaks down.
Then, the difference between the site and link solitons becomes remakable.
Specifically, in the region of $\xi \ll d$, a small fluctuation from the site soliton decreases the total energy of the system because the presence of a density hole at a site costs a significant amount of the interaction energy. 
This indicates that $\hbar \omega_1/J$ for the site soliton should take a large negative value at large $\kappa$ and this is the case realized in Fig.~\ref{3031_bogo_uni}(c). 
By contrast, a small fluctuation from the link soliton increases the total energy because it digs a hole in the central sites. 
This indicates that $\hbar \omega_1/J$ for the link soliton should take a large positive value at large $\kappa$. 
However, since $\hbar \omega_1/J$ is negative in the small-$\kappa$ region, $\hbar \omega_1/J$ has to touch zero at a certain $\kappa$, which is nothing but $\kappa_c$, before the sign change. 
The touch on zero of $\hbar \omega_1/J$ results in the coupling to the anti-mode of itself, leading to the emergence of the dynamical instability. 
The dynamical instability caused by a similar mechanism has been seen in a dark soliton of a 1D BEC pinned by a repulsive potential barrier \cite{ichihara_matter-wave_2008}.

In the following section, we will analyze the stability of the dark soliton within the semiclassical regime, focusing on the region of $\kappa_c \leq \kappa \leq 0.7$ where the dark soliton is dynamically unstable only if $M$ is even.

\subsection{Weakly quantum regime} \label{ss4_sssb_box_weak}

\begin{figure*}
    \centering
    \includegraphics[width=0.55\linewidth] {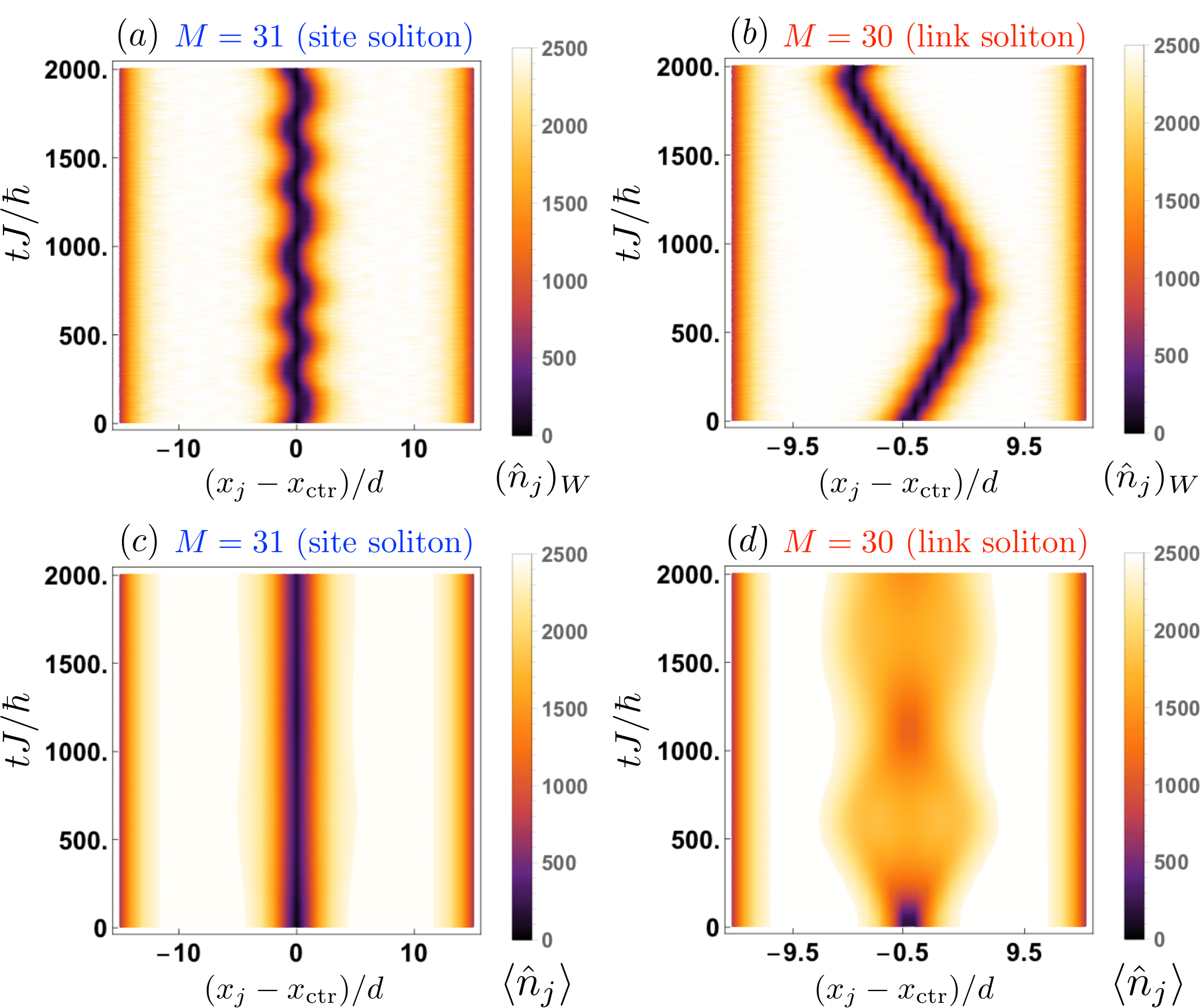}
    \vspace{-2mm}
    \caption{Time evolution of the density profile in the system with a box potential for $\gamma = 6.25 \times 10^{-8}$, $\kappa=0.5$ and $\nu=2000$. (a) and (b) represent the density profile in a single shot for $M=30$ and $31$. (c) and (d) represent  the density profile averaged over all the TWA samples, corresponding to its quantum average, for $M=30$ and $31$.}
    \label{nuj0.5_nu2000}
\end{figure*}
    
\begin{figure*}
        \centering
        \includegraphics[width=0.55\linewidth] {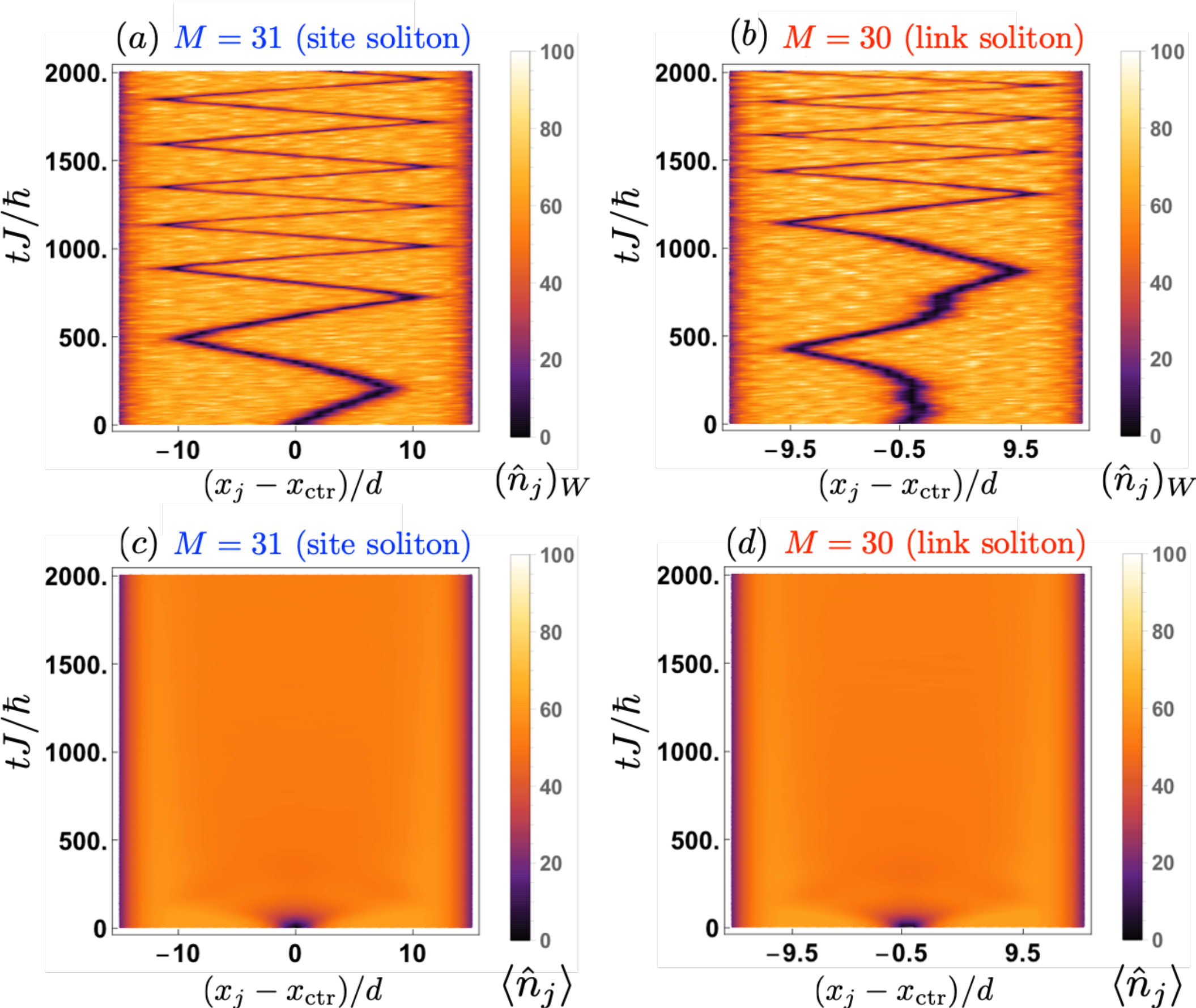}
        \vspace{-2mm}
        \caption{Time evolution of the density profile in the system with a box potential for $\gamma = 1.0 \times 10^{-4}$, $\kappa=0.5$ and $\nu=50$. (a) and (b) represent the density profile in a single shot for $M=30$ and $31$. (c) and (d) represent the quantum average of the density profile for $M=30$ and $31$.}
        \label{nuj0.5_nu50}
    \end{figure*}

In this section, we analyze the semiclassical dynamics of a dark soliton within the TWA.
According to the previous studies using the MPS method, a dark soliton is dynamically unstable in a regime of strong quantum fluctuations, even if the soliton is dynamically stable in the classical regime~\cite{mishmash_quantum_2009, mishmash_quantum_2009-2, mishmash_ultracold_2009}.
This indicates that the dynamical stability of a dark soliton changes between the two regimes.
In order to corroborate the switch of the dynamical stability due to quantum fluctuations, we perform TWA analyses of soliton dynamics for a wide range of $\gamma\equiv U/(z \nu J)$, which controls the strength of quantum fluctuations, while fixing the value of $\kappa \equiv \nu U/J$.
Specifically, in the case of a box potential, we set the nonlinearity strength as $\kappa = 0.5$, at which the site soliton for $M=31$ is dynamically stable while the link soliton for $M=30$ is dynamically unstable.
It is worth noting that in the regime of weak quantum fluctuations that is treated in the present work the average filling factor $\nu$ is as large as $30 \leq \nu \leq 5000$ and $\kappa \leq 0.7$. 
In such a parameter region, it is practically impossible to compute the soliton dynamics by means of the MPS method because the size of the local Hilbert space is too large.

In Fig.~\ref{nuj0.5_nu2000}, we show the time evolution of the density profile for $\nu=2000$, where quantum fluctuations are weak. 
The initial state for the time evolution possesses a standing dark soliton at the center of the system.
Figures~\ref{nuj0.5_nu2000}(a) and (b) represent dynamics in a single shot arbitrarily chosen among the TWA samples for $M=31$ and $30$.
In Fig.~\ref{nuj0.5_nu2000}(a), we see that the site soliton remains localized near the center during the time evolution, i.e., it is dynamically stable.
In Fig.~\ref{nuj0.5_nu2000}(b), we see that the link soliton departs away from its initial position in an early stage and moves over a broad range of the system in a later stage, i.e., it is dynamically unstable. 
These results are consistent with the classical linear stability analysis presented in Sec.~\ref{ss4_sssa_box_cla}.
Figures~\ref{nuj0.5_nu2000}(c) and (d) represent the time evolution of the quantum average of the density profile, $\langle \hat{n}_j (t)\rangle$.
While the behavior of the site soliton is almost identical to the single shot (compare Fig.~\ref{nuj0.5_nu2000}(c) with (a)), the notch in the average density of the link solition becomes filled as time evolves in contrast to the single-shot result (compare Fig.~\ref{nuj0.5_nu2000}(d) with (b)). 
This happens because the link soliton moves depending on a random initial velocity in each single shot and averaging such random motions of the density notch leads to the apparent disappearance of the notch~\cite{delande_many-body_2014, dziarmaga_images_2003, dziarmaga_quantum_2004}.

In Fig.~\ref{nuj0.5_nu50}, we show the time evolution of the density profile for stronger quantum fluctuations, say $\nu=50$. 
In a single shot shown in Figs.~\ref{nuj0.5_nu50}(a) and (b), we see that for both cases of $M=31$ and $30$ the dark soliton departs away from its initial position in an early stage and moves over a broad range of the system in a later-time stage. 
In the quantum average, the density notch of the dark soliton becomes filled in a short time. 
Thus, in the regime of relatively strong quantum fluctuations, both of the site and link solitons are dynamically unstable in contrast to the results of the classical stability analysis.

In order to quantify the change in the stability of the dark soliton, we calculate the decay time $\tau_c$~\cite{mishmash_quantum_2009-2}, which can be extracted from the soliton contrast,
\begin{equation}
C(t) \equiv \frac{\langle \hat{n}_{\rm{max}}\rangle -\langle \hat{n}_{\rm{mid}}\rangle}{\langle \hat{n}_{\rm{max}}\rangle +\langle \hat{n}_{\rm{mid}}\rangle},
\end{equation}
where $\langle \hat{n}_{\rm{max}}\rangle$ is the maximal averaged density over all sites $j$ and $\langle \hat{n}_{\rm {mid}}\rangle$ is the averaged density at the center of the system. 
At the initial time, $t=0$, the averaged density at the center satisfies $\langle \hat{n}_{\rm{mid}}(0)\rangle \simeq 0$ because the dark soliton is located at the center of the system.
If the notch of the averaged density decays over time, the soliton contrast approaches $C \simeq 0$ from $C(0) \simeq 1$. 
We define the decay time $\tau_c$ as the half-life of $C$ from the initial value $C(0)$,
\begin{equation}
\frac{C(\tau_c)}{C(0)}=\frac{1}{2}.
\end{equation}

\begin{figure}
    \centering
    \includegraphics[width=0.9\linewidth] {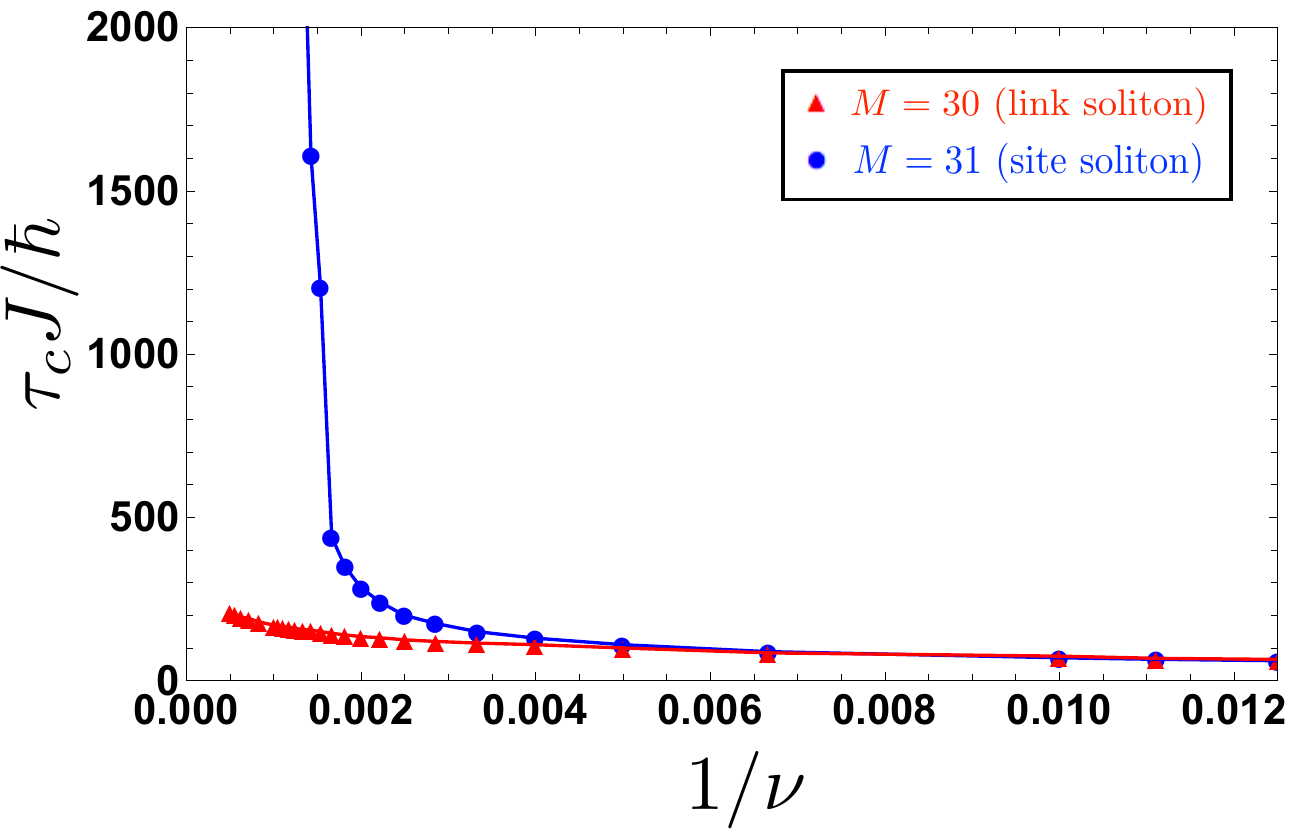}
    \caption{Classical-to-quantum crossover of the stability of the two types of dark soliton in a box potential by quantum fluctuations for $\kappa=0.5$. We show the dependence of the decay time $\tau_c J/ \hbar$ on quantum-fluctuation strength $1/\nu$ for $M=31$ and $M=30$. When $1/\nu$ increases, $\tau_c J/ \hbar$ of the site soliton approaches that of the link soliton.}
    \label{dt}
    \end{figure}%

In Fig.~\ref{dt}, we show the decay time $\tau_c J/\hbar$ versus $1/\nu$ for $M=30$ (red triangles) and $31$ (blue circles). 
Here $1/\nu$ is proportional to $\sqrt{\gamma}$, which characterizes the strength of quantum fluctuations, because we fix $\kappa$.
In a regime of weak quantum fluctuations, e.g., $1/\nu\leq 0.001$, $\tau_c J/\hbar$ for $M=31$ is much larger than that for $M=30$, reflecting the difference between the classical stabilities of the site and link solitons. 
When $1/\nu$ increases, the former $\tau_c J/\hbar$ approaches the latter. 
In the region of $1/\nu \geq 0.005$, both $\tau_c J/\hbar$ coincide with each other. 
Thus, by calculating $\tau_c J/\hbar$ versus $1/\nu$ we have successfully identified the crossover behavior of the dark-soliton stability from the classical regime to the quantum regime.

\begin{figure}
    \centering
    \includegraphics[width=0.9\linewidth] {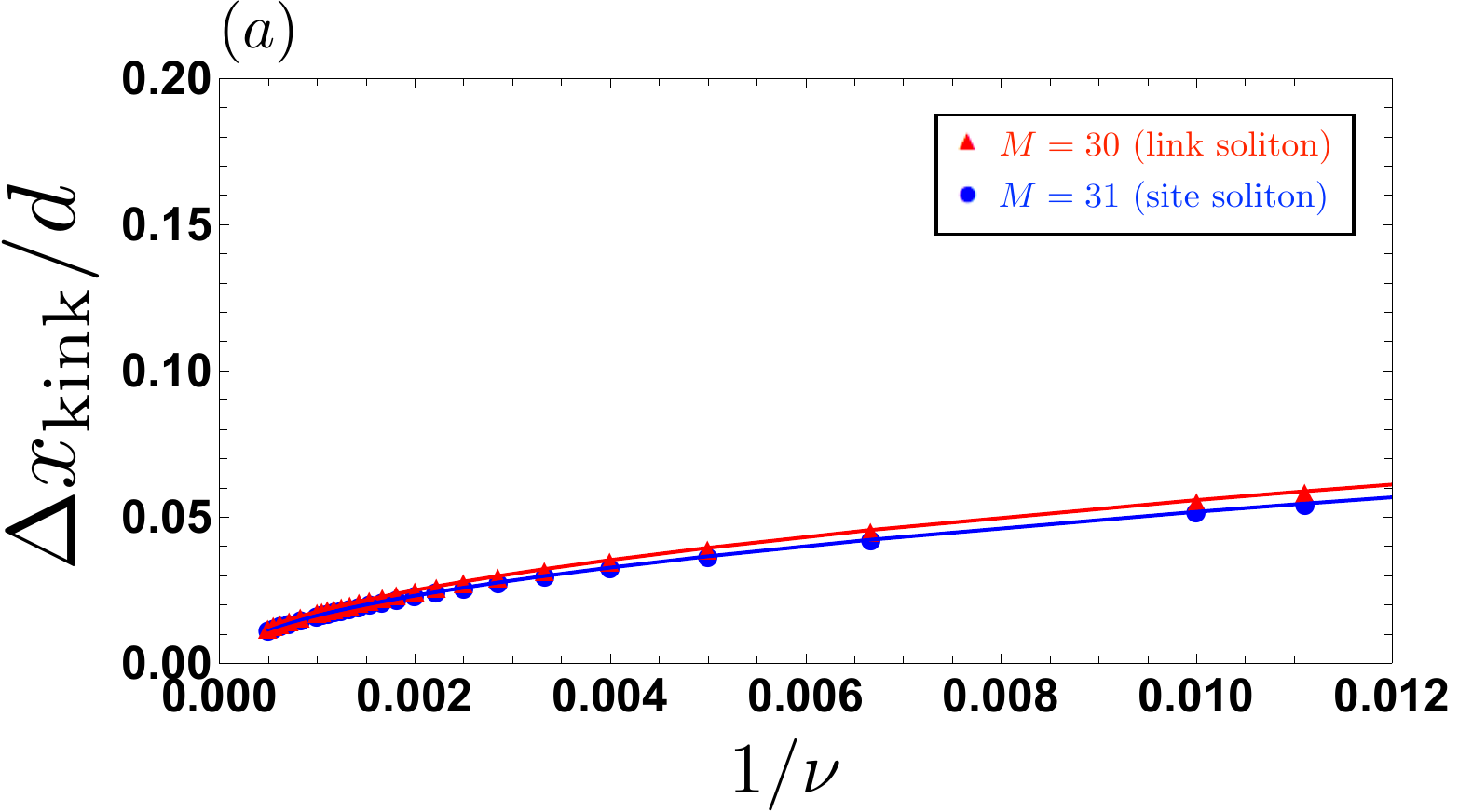}
    \includegraphics[width=0.9\linewidth] {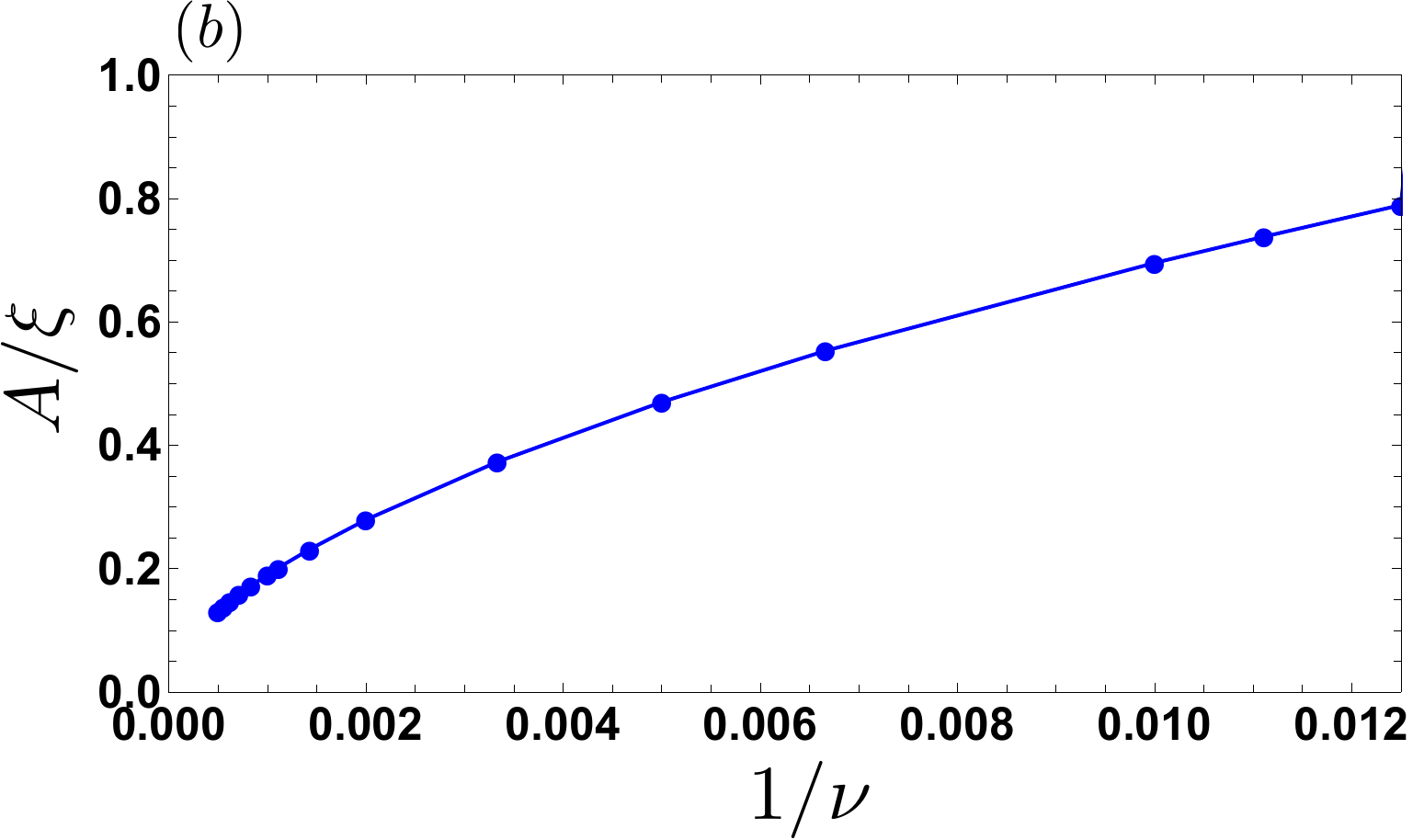}
    \caption{The two length scales related to the effect of quantum fluctuations on the dark soliton for $\kappa = 0.5$. (a) The standard deviation of the initial position of the density notch $\Delta x_{\rm kink}$ in unit of the lattice constant $d$ versus the fluctuation strength $1/\nu$ for $M=31$ (blue circles) and $M=30$ (red triangles). (b) The oscillation amplitude estimated by the initial velocity $A$, estimated by the initial velocity, in unit of the healing length $\xi$ versus $1/\nu$ for $M=31$.}
    \label{dt_fac}
    \end{figure}%

The dynamical instability due to quantum fluctuations can be attributed to a large-amplitude oscillation induced by the initial random noise present in the TWA calculations. 
In order to corroborate this scenario, we show in Fig.~\ref{dt_fac} (a) the standard deviation of the initial position of the soliton notch $\Delta x_{\rm kink}/d$ and (b) the averaged amplitude of soliton oscillations during the time evolution $A/\xi$ as functions of $1/\nu$. 
On the one hand, in Fig.~\ref{dt_fac} (a), we see that $\Delta x_{\rm kink}$ is much smaller than $d$ over the parameter region, in which the crossover happens. 
This means that the fluctuation in the initial position of the soliton notch is not a main cause for the quantum dynamical instability. 
On the other hand, the amplitude $A$ increases with increasing $1/\nu$ to become comparable to $\xi$ in the crossover regime. 
Since the classical linear stability analysis is valid under the assumption that amplitude  of normal modes is small, the emergence of the oscillation with the amplitude comparable with $\xi$ implies the breakdown of the classical picture.


\section{System with a parabolic trap potential} \label{ss5_para_results}
In this section, we consider a situation in which a 1D lattice Bose gas is confined in a parabolic trap potential. 
In our formulation, this corresponds to setting $\epsilon_j$ in the BHH (\ref{eq:BHH}) as
\begin{eqnarray}
\epsilon_j=\Omega (j-x_{\rm ctr}/d)^2.
\end{eqnarray}
The coefficient $\Omega$ is related to the trap frequency $\omega_z$ as $\Omega = \frac{1}{2}m\omega_z^2 d^2$, where $m$ denotes the bare mass of the boson. 
Since a parabolic trap potential is present in typical ultracold-gas experiments, it is important to investigate whether or not our findings for a box potential shown in Sec.~\ref{ss4_box_results}, especially the classical-to-quantum crossover of the soliton stability, are relevant even in the presence of the trap potential.

For a systematic analysis of quantum-fluctuation effects on the soliton stability, we need to control the size of the gas $R$ and the filling factor at the center of the system $\nu$.
Notice that we define $R$ and $\nu$ as those of the ground state because $\nu$ in the dark soliton state is nearly equal to zero.
In the case of a box potential, $R$ is simply equal to $L$ and $\nu$ is approximately given by $\nu = N/M$ as long as $\xi \ll L$, where $N$ is the total number of particles. 
By contrast, in the presence of a parabolic potential, $R$ and $\nu$ significantly depend not only on $N$ but also on $\Omega/U$. 
Specifically, when the nonlinearity of the DNLSE is strong enough for the condition $R \gg a_{z}$ to be satisfied, the Thomas-Fermi approximation provides the following analytical expression \cite{pitajevskii_bose-einstein_2003, rey_ultracold_2004},
\begin{eqnarray}
    R = 2\left(\frac{3UN}{4\Omega} \right)^{\frac{1}{3}}d,
    \label{eq:TFR}
    \\
    \nu_{\rm TF} = \left(\frac{9N^2\Omega}{16U}\right)^{\frac{1}{3}},
    \label{eq:TFnu}
\end{eqnarray}
where $a_{z} = d(J/\Omega)^{\frac{1}{4}}$ denotes the harmonic oscillator length and $\nu_{\rm TF}$ means the filling factor at the trap center within the Thomas-Fermi approximation. 
The length scale $R_{\rm TF} = R/2$ is often called the Thomas-Fermi radius. 
Using Eqs.~(\ref{eq:TFR}) and (\ref{eq:TFnu}), we adjust the three parameters $N$, $\Omega/J$ and $U/J$ in such a way that $R$, $\nu_{\rm TF}$, and $\kappa_{\rm TF}\equiv \nu_{\rm TF} U/J$ become desired values. 
In the calculations shown below, we set $R = 30$. 
Moreover, we set $M=51$ ($M=50$) for the case of the site (link) soliton.

\subsection{Classical regime} \label{ss5_sssa_para_cla}

\begin{figure}
    \centering
    \includegraphics[width=0.9\linewidth] {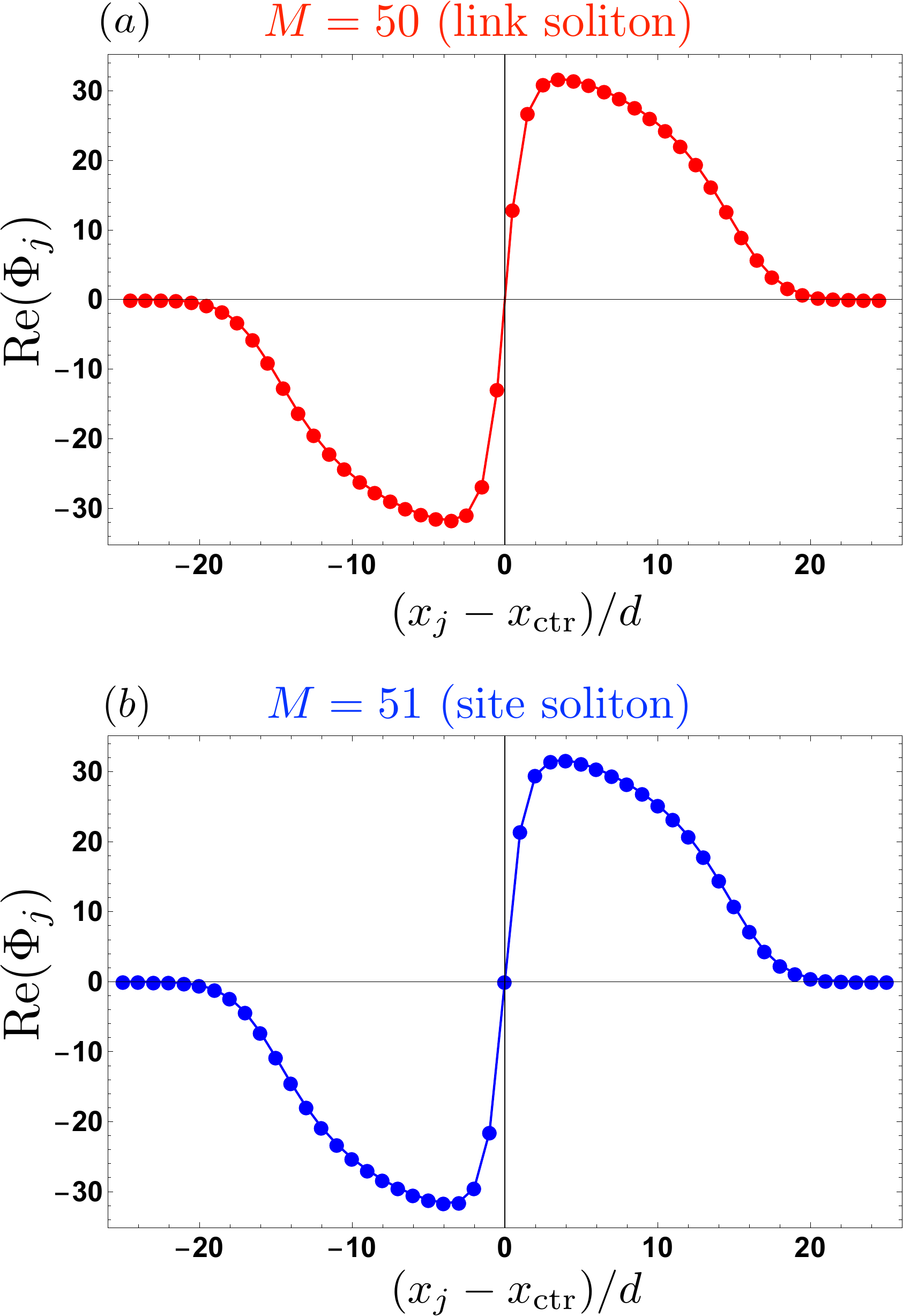}
    \caption{The spatial profile of the stationary solutions $\Phi_j$ at $\kappa_{\rm TF}= 1.0$ and $\nu_{\rm TF}=1000$ for (a)$M=50$ and (b)$M=51$. The solid lines are guide to the eye.}
    \label{har_sta_nu1000}
    \end{figure}
\begin{figure}
    \includegraphics[width=0.9\linewidth] {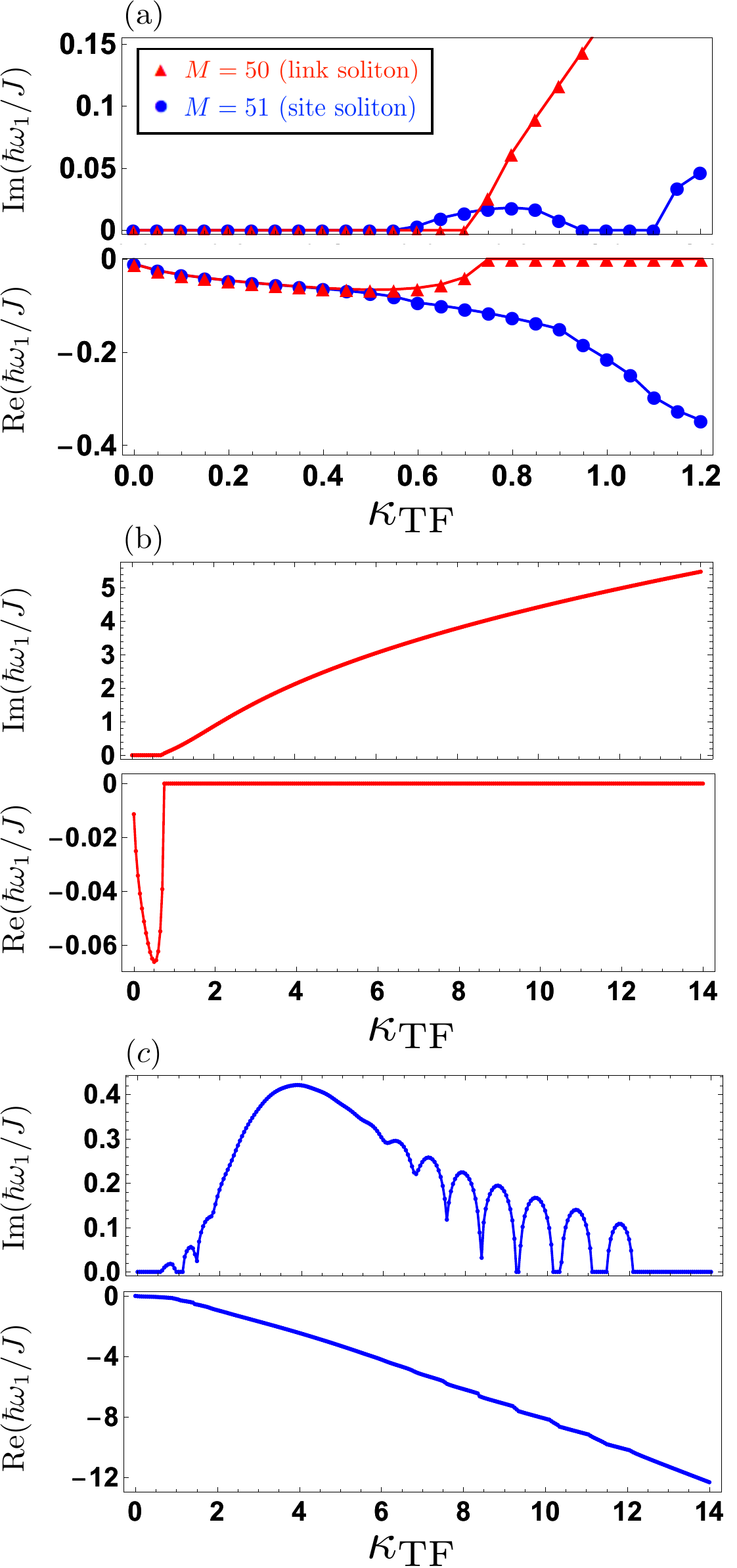}
    \vspace{-1mm}
    \caption{Frequency $\hbar \omega_1 / J$ of the normal mode that is relevant to the energetic and dynamical instability of the dark soliton as a function of $\kappa_{\rm TF}$, where $M=50$ (red triangles) and $51$ (blue circles). (a) displays $\hbar \omega_1 / J$ in the small-$\kappa_{\rm TF}$ region, $0\leq \kappa \leq 1.2$, for both $M=50$ and $51$. (b) and (c) display $\hbar \omega_1 / J$ in the expanded region, $0\leq \kappa_{\rm TF} \leq 14$, for $M=50$ and $51$, respectively. }
    \label{har_bogo}
    \end{figure}

%
Figure \ref{har_sta_nu1000} depicts the spatial profile of $\Phi_j$ of the dark-soliton solutions at $\kappa_{\rm TF} = 1.0$ for $M=50$ and $51$ in the presence of a parabolic trap potential. 
We perform the classical linear stability analysis of such soliton solutions on the basis of the Bogoliubov equations for various values of $\kappa_{\rm TF}$.
Figure \ref{har_bogo} depicts the frequency $\hbar \omega_1 / J$ of the normal mode that is relevant to the energetic and dynamical instabilities of the dark soliton as a function of $\kappa_{\rm TF}$.

 We see in Fig.~\ref{har_bogo}(a) that for $\kappa_{\rm TF} \lesssim 0.3$, $|\hbar \omega_1 / J|$ increases as $\kappa_{\rm TF}$ increases. 
 This is in stark contrast to the case of a box potential shown in Fig.~\ref{3031_bogo_uni}(a). 
 The reason for this behavior is rather simple. 
 When $\kappa_{\rm TF}$ increases for fixed $R$ and $\nu_{\rm TF}$, $\Omega/J$ increases. 
 Since $\hbar \omega_1 / J \simeq -\sqrt{2\Omega/J}$ in the continuum limit~\cite{muryshev_stability_1999}, the value $|\hbar \omega_1 / J|$ increases with $\Omega/J$. 
The tendency of $\hbar \omega_1 / J$ in a region of larger $\kappa_{\rm TF}$ is similar to the case of a box potential. 
On the one hand, when $\kappa_{\rm TF}$ exceeds a certain critical value ($\kappa_{\rm TF}\simeq 0.75$), ${\rm Re}(\hbar \omega_1 / J)$ vanishes and ${\rm Im}(\hbar \omega_1 / J)$ becomes finite for $M=50$, signaling the dynamical instability of the link soliton. 
On the other hand, when $\kappa_{\rm TF}$ exceeds another certain critical value ($\kappa_{\rm TF}\simeq 0.6$), ${\rm Re}(\hbar \omega_1 / J)$ keeps growing but ${\rm Im}(\hbar \omega_1 / J)$ becomes finite for $M=51$ due to the coupling with anti-modes, signaling the dynamical instability of the site soliton. 
However, we emphasize that there is the parameter region $0.95 \leq \kappa_{\rm TF} \leq 1.1$, in which the site soliton is dynamically stable while the link soliton is dynamically unstable as is the case analyzed in Sec.~\ref{ss4_box_results}.

\subsection{Weakly quantum regime} \label{ss5_sssb_para_weak}
In this section, we present results of semiclassical dynamics of a dark soliton in a parabolic potential obtained by using the TWA. 
Since our main purpose here is to corroborate the relevance of the classical-to-quantum crossover, which we found for a box potential, to the case of the parabolic potential, we set $\kappa_{\rm TF} = 1.0$, at which the site (link) soliton is dynamically stable (unstable) in the classical limit.

\begin{figure*}
\centering
\includegraphics[width=0.55\linewidth] {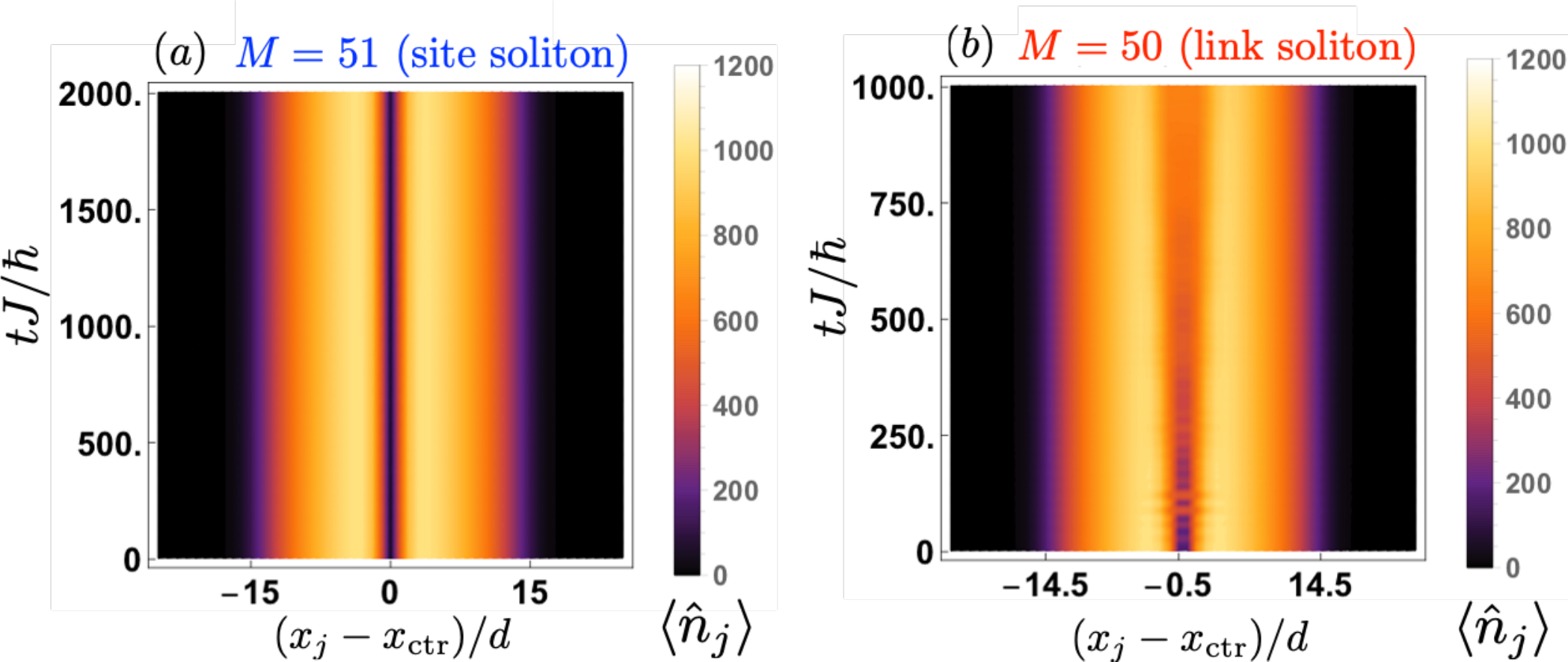}
\vspace{-2mm}
\caption{Time evolution of the density profile in the system with a parabolic trap potential for $\kappa_{\rm TF}=1.0$ and $\nu_{\rm TF}=1000$. (a) and (b) represent the density profile averaged over all the TWA samples, corresponding to its quantum average, for $M=50$ and $51$.}
\label{har_nuj1.0_nu1000}
\end{figure*}%

Figure~\ref{har_nuj1.0_nu1000} depicts the time evolution of the average density profile, $\langle \hat{n}_j(t) \rangle$, for (a) $M=51$ and (b) $M=50$. 
The initial state for the time evolution possesses a standing dark soliton at the center of the system. 
There we set $\nu=1000$, at which quantum fluctuations are rather weak.
We see that the site soliton for $M=51$ shown in Fig.~\ref{har_nuj1.0_nu1000}(a) is dynamically stable while the density notch of the link soliton for $M=50$ in Fig.~\ref{har_nuj1.0_nu1000}(b) decays in time.
In short, the stability of the dark solitons in this regime of very weak quantum fluctuations is consistent with the classical stability as is the case in the system with a box potential.

\begin{figure*}
\centering
\includegraphics[width=0.55\linewidth] {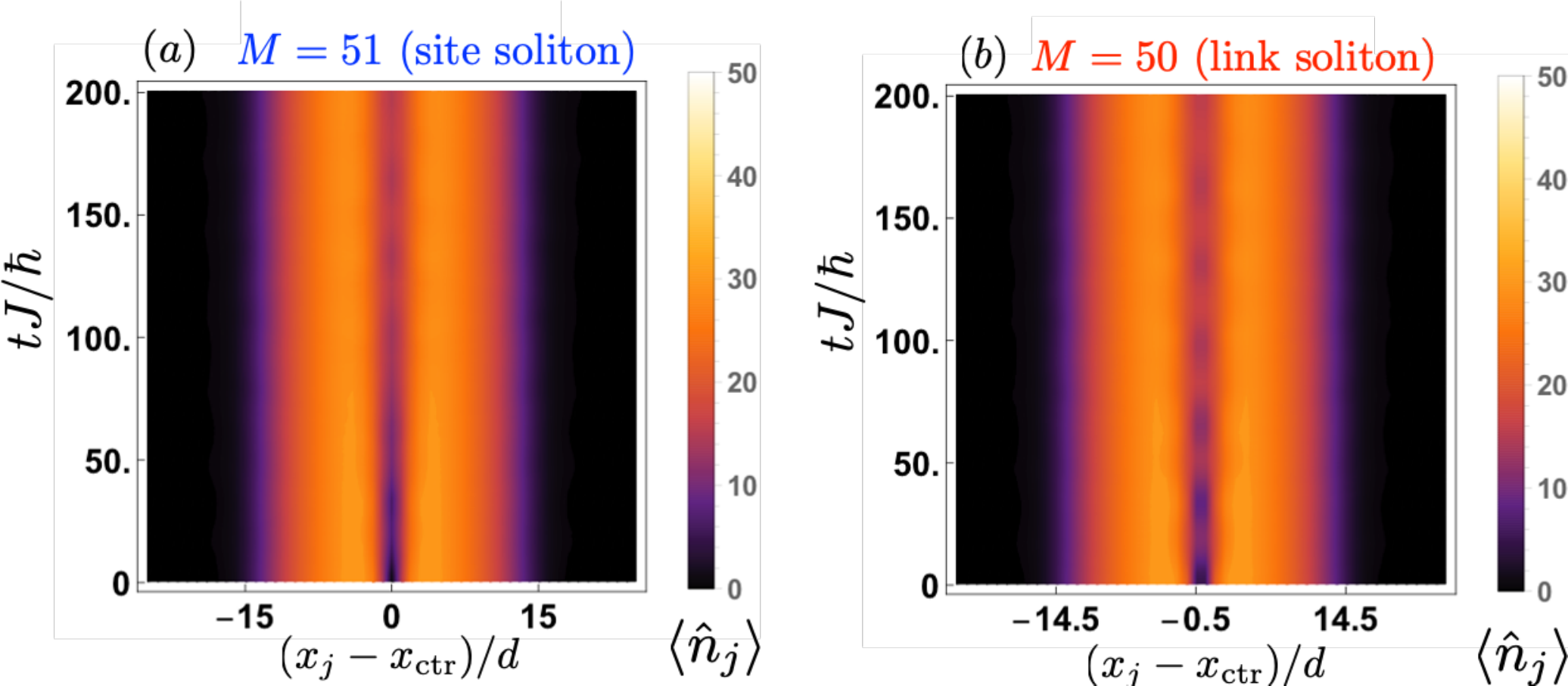}
\vspace{-2mm}
\caption{{Time evolution of the density profile in the system with a parabolic trap potential for $\kappa_{\rm TF}=1.0$ and $\nu_{\rm TF}=30$.} {(a) and (b) represent the density profile averaged over all the TWA samples, corresponding to its quantum average, for $M=50$ and $51$.}}
\label{har_nuj1.0_nu30}
\end{figure*}%

In Fig.~\ref{har_nuj1.0_nu30}, we take $\nu = 30$, at which quantum fluctuations are relatively strong, and show the time evolution of the average density profile for (a) $M=51$ and (b) $M=50$.
The density notches of both site and link solitons decay in time, meaning that both are dynamically unstable. 
Since the site soliton is dynamically stable in the classical limit, its instability should be due to quantum fluctuations.

\begin{figure}
\centering
\includegraphics[width=0.9\linewidth] {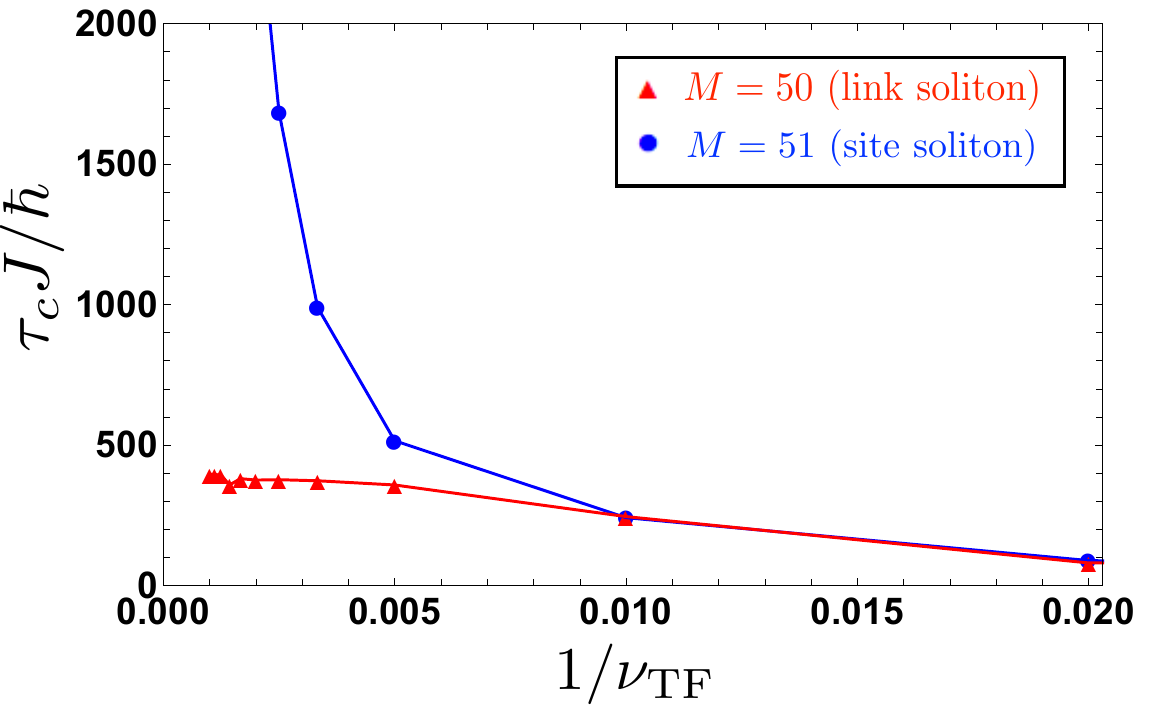}
\vspace{-2mm}
\caption{Classilal-to-quantum crossover of the stability of the two types of dark soliton in a parabolic trap potential by quantum fluctuations for $\kappa_{\rm TF}=1.0$. We show the dependence of the decay time $\tau_c J/\hbar$ on quantum-fluctuation strength $1/\nu_{\rm TF}$ for $M=51$ and $M=50$. When $1/\nu_{\rm TF}$ increases, $\tau_c J/\hbar$ of the site soliton approaches that of the link soliton in both potentials.}
\label{har_nuj1.0_ds}
\end{figure}

In order to quantify the classical-to-quantum crossover of the soliton stability, we show in Fig.~\ref{har_nuj1.0_ds} the decay time of the soliton contrast $\tau_c J/\hbar$ as a function of $1/\nu_{\rm TF}$. 
In a regime of very weak quantum fluctuations, say $1/\nu_{\rm TF} \leq 0.002$, $\tau_c J/\hbar$ for the site soliton is much larger than that for the link soliton, corresponding to the classical regime. 
When $1/\nu_{\rm TF}$ increases, $\tau_c J/\hbar$ for the site soliton approaches that for the link soliton. 
They almost coincides with each other when $1/\nu_{\rm TF} \geq 0.01$.
Thus, the system with a parabolic trap potential exhibits the classical-to-quantum crossover with respect to dynamical stability of the two types of dark soliton, as seen in the system with a box potential.


\section{Conclusions} \label{ss6_conclusion}
In conclusion,  we have investigated the stability of a dark soliton in a one-dimensional lattice Bose gas, focusing on effects of weak quantum fluctuations on it.
First, we revisited the stability analysis in the classical regime on the basis of the Gross-Pitaevskii mean-field theory to clarify reasons why the stability of a dark soliton differs depending on whether its phase kink is located at a lattice site or a site at two sites.
The link soliton is dynamically unstable when the nonlinear parameter $\kappa = \nu U/J$ is larger than a critical value.
By contrast, there are some range of $\kappa$ in which the site soliton is dynamically stable above the critical value.
The dependence of the instability results from the difference of energy gain by small oscillations, which is determined by the competition of three length scales, the box size $M$, the healing length $\xi$ and the lattice constant $d$.
Although the dynamical instability of the link soliton can be explained by this mechanism, we reported the emergence of the dynamical instability for the site soliton due to a different mechanism for the large $\kappa$ region.

To elucidate the effects of weak quantum fluctuations on the soliton stability, we analyzed the time-evolution of a dark soliton under a box potential and a harmonic potential within the truncated Wigner approximation.
We numerically showed that when the strength of quantum fluctuation increases, the lifetime of a site soliton gradually approaches that of a link soliton in both the potentials.
This means that the distinction in the two-types of dark soliton, which is present in the classical limit, is smeared out by quantum fluctuations.
 This finding enables us to distinguish the instability of a dark soliton due to quantum fluctuations from the classical dynamical instability by observing the difference in dynamics between the two types of dark solitons.
We also discussed a mechanism for quantum fluctuations to destabilize the dark soliton and revealed that quantum fluctuations amplify the oscillation of the site soliton, which leads to its destabilization.


\acknowledgements
The authors thank Masaya Kunimi and Shimpei Goto for fruitful comments and discussions.
This work was supported by JSPS KAKENHI Grants No.~18K03492 (I.D.), No.~18H05228 (I.D.), and No.~18K03472 (K.K.), by a research grant from CREST, JST No.~JPMJCR1673, and by the Q-LEAP program of MEXT, Japan No.~JPMXS0118069021.



%


\end{document}